\documentclass[aps,pra,twocolumn,a4paper]{revtex4-1}
\usepackage{amsmath,amssymb,epsfig}
\usepackage{color}
\usepackage{float}
\usepackage{enumerate}
\usepackage{soul}
\usepackage{graphicx}
\usepackage{hyperref}
\usepackage[utf8]{inputenc}
\usepackage{lineno}

\newcommand{\ket}[1]{ \left|#1\right>}

\newcommand{\expect}[1]{ \left<#1\right>}

\begin{document}

\title{{\Large{\bf Lasing by driven atoms-cavity system in collective strong coupling regime}}}
\author{Rahul Sawant}
\email{rahuls@rri.res.in}
\author{S. A. Rangwala}
\email{sarangwala@rri.res.in}
\affiliation{Light and Matter Physics Group, Raman Research Institute, Sadashivanagar, Bangalore 560080, India}

\begin{abstract}
The interaction of laser cooled and trapped atoms with resonant light is limited by the linewidth of the excited state of the atom. Another precise optical oscillator is an optical Fabry-P\'erot cavity. The combining of cold atoms with optical oscillators is emerging as an area with great potential for precision measurements and the creation of versatile quantum optics systems. Here we show that when driven atoms are in the collectively strongly coupled regime with the cavity, exhibiting  vacuum Rabi splitting (VRS), lasing is observed for the emitted light, red detuned from atomic transition. This is demonstrated experimentally by the observation of a lasing threshold, polarisation purity, mode purity, and line narrowing. The laser is created spontaneously by the atomic emission into the cavity mode, which stimulates cavity emission, and is capable of operating continuously without a seed laser. The gain mechanism is understood by theoretical modelling and illustrates why the observed lasing is generic to the coupled system. This opens up a range of possibilities of using the phenomenon for a variety of new measurements. 
\end{abstract}

\maketitle

\section{Introduction}
While both cold atoms~\cite{ketterle_nobel_2002,cornell_nobel_2002}and cavity physics~\cite{thompson_observation_1992,raimond_manipulating_2001} are central to atomic physics, few experiments combine an intra-cavity ensemble of cold atoms with a cavity~\cite{chan_observation_2003,hernandez_vacuum_2007,guerin_mechanisms_2008,
baumann_dicke_2010,ray_temperature_2013,purdy_tunable_2010}. Cooled and trapped atoms are spatially localized, natural linewidth limited and have high local densities, which permits ease of spectral and spatial overlap with cavity modes. When a large number of these atoms are contained within the cavity mode volume, and the atom-cavity system is brought into resonance, a collective strong coupling of atoms and light alters~\cite{tavis_exact_1968,raizen_normal-mode_1989,thompson_observation_1992,hernandez_vacuum_2007,ray_temperature_2013,sharma_optical_2015,sawant_optical-bistability-enabled_2016,dutta_nondestructive_2016} the transmission properties of resonant probe light through the cavity. This manifests in a change from a single transmission peak per spatial electromagnetic (EM) mode with a Lorentzian lineshape for an empty cavity, to zero transmission of the probe on atomic resonance and on scanning the probe light around the atomic transition, symmetric red and blue detuned transmission peaks are observed. The frequency difference between the transmitted peaks is given as $g_0 \sqrt{N_c}$, where $g_0$ is the single atom-cavity coupling and $N_c$ is the effective number of atoms coupled to the cavity mode. 

In this article, we trap within the cavity mode, fluorescing and continuously driven MOT atoms. The atom numbers in the present experiment are such that the collective strong coupling regime of atom-cavity mode interaction is accessible. We then pose the question, what is the effect of externally driven atoms on the coupled atom-cavity system? The answer is surprisingly different from the non-driven system. On scanning the cavity length, the system now emits light out of the cavity at red and blue detuning of the cavity with respect to the atomic resonance frequency. Since the magento-optical trap (MOT) lasers are also the drive lasers and are red detuned, the experimentally broken symmetry reflects in the emission of light from the cavity atoms in a dramatic manner, where the red detuned light emitted is a laser and the blue detuned light is not a laser. Evidence of lasing for the red detuning is obtained from the series of experiments discussed below. The lasing here is not due to population inversion in the gain medium but due to multi-photon processes as is discussed in the theoretical analysis. The lasing is seeded by spontaneous emission from the driven atoms and not by an external seed/probe laser. When this system is probed with an external cavity probe laser, Fano-like resonances and line narrowing is seen, which is much narrower than all the line widths that are natural to the system, providing irrefutable evidence of lasing by the driven atom-cavity system. 

In recent times, understanding lasing with different gain mechanisms ~\cite{mompart_lasing_2000,scully_quantum_1997,vrijsen_raman_2011,
hilico_operation_1992,guerin_mechanisms_2008,kruse_observation_2003,
slama_superradiant_2007,mckeever_experimental_2003,bohnet_steady-state_2012,norcia_cold-strontium_2016} has accelerated. For example, one such process relies on quantum interference between probability amplitudes in multilevel atoms and does not require population inversion~\cite{mompart_lasing_2000, scully_quantum_1997}. Raman processes between different hyperfine levels and sublevels~\cite{vrijsen_raman_2011,hilico_operation_1992,guerin_mechanisms_2008}, levels involving external degrees of freedom~\cite{kruse_observation_2003,slama_superradiant_2007} have been exploited as gain mechanisms including laser involving a single atom~\cite{mckeever_experimental_2003}. Much progress in understanding the role of collective effects on lasing~\cite{slama_superradiant_2007,bohnet_steady-state_2012,norcia_cold-strontium_2016} has been made. William \textit{et.al}~\cite{guerin_mechanisms_2008} showed lasing using three different gain mechanisms, Mollow gain, Raman transition between Zeeman sublevels and gain due to four-wave mixing, all of which were achieved with the same gain medium. Raman transition between Zeeman sublevels as a gain process was used by the same group to demonstrate random laser without any cavity and seed laser~\cite{baudouin_cold-atom_2013}.
In the experiments here we demonstrate that continuous lasing triggered by spontaneous emission by the driven, non-inverted population of atoms in the cavity mode is possible, robust and natural under reasonable experimental conditions. The gain mechanism for the lasing action is a result of multi-photon scattering~\cite{grynberg_central_1993, wu_observation_1977}. However, unlike previous works~\cite{guerin_mechanisms_2008, wu_observation_1977}, we observe lasing action even when the cavity is not driven by a seed laser. Lezama \textit{et.al}~\cite{lezama_radiative_1990} have earlier seen lasing in an atomic beam experiment, but the atom-cavity coupling was very small for any co-operative effect to manifest.


%
\onecolumngrid

\begin{figure}[h]
\centering
\includegraphics[width=12cm]{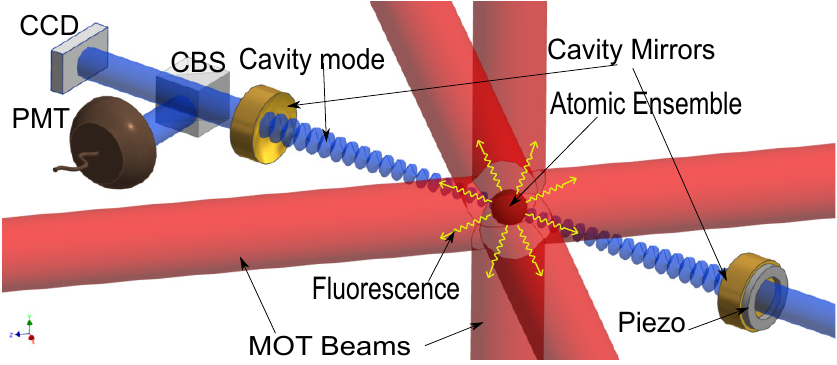}
\caption{Experimental schematics. $^{85}$Rb atoms are cooled by the six red MOT beams and trapped with the help of magnetic gradient. The centre of MOT beams and centre of the cavity formed by the two mirrors overlap. PMT - Photo multiplier tube, CBS - Common beam splitter, CCD- Charge coupled device (camera).}
\label{exp}
\end{figure}
\twocolumngrid
\section{The experiment}

\begin{figure}[H]
\centering
\includegraphics[width=15cm]{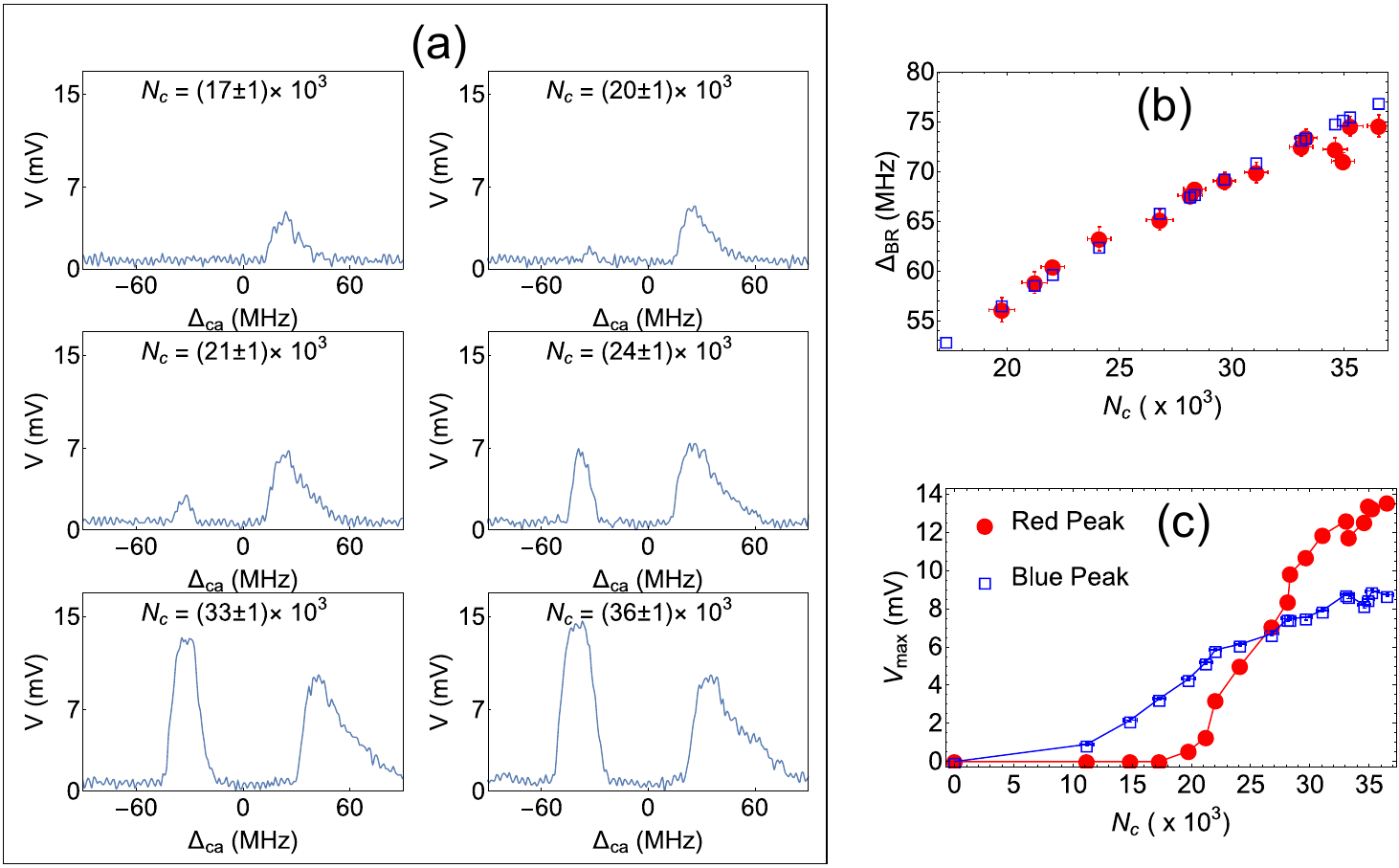}
\caption{Change in emission from the cavity when the atom number coupled to the cavity is varied. (a) These graphs show cavity emission spectrum for different atom numbers. $\Delta_{ca} = \omega_c - \omega_a $ is detuning of empty cavity resonance from atomic transition and V is signal detected on PMT. (b) Red dots with error bars show the change in peak separation of the cavity emission with $N_c$. Blue empty squares are corresponding VRS ($2g_0 \sqrt{N_c}$) values. $\Delta_{BR}$ - The frequency separation between red and blue detuned peaks (c) Variation in the height of red and blue detuned peaks. In the range of measurement, the red peak shows a threshold behaviour with atom number whereas the blue peak does not. $V_{max}$ - Maximum voltage generated by PMT for the red and blue detuned peaks. Total intensity of all cooling laser beams is  36 mW/cm$^2$ for all the measurements.}
\label{atnogrid}
\end{figure}

The experimental schematic is as shown in the Fig.~\ref{exp}, $^{85}$Rb atoms are cooled and trapped in a Magnet-optical trap (MOT). The centre of the MOT and centre of the Fabry-P\'erot cavity formed around it are made to overlap~\cite{ray_thin_2014, ray_temperature_2013}. The atomic ensemble is constantly illuminated by the six MOT light beams. There is no light incident on the cavity mirrors. The atom-cavity is in collective strong coupling regime, i.e., $g_0 \sqrt{N_c} > \Gamma, \kappa$~\cite{thompson_observation_1992,agarwal_spectroscopy_1998}. Here $g_0/2\pi = 201$ kHz is maximum single atom coupling strength, $\Gamma/2\pi =  6.06$ MHz is atomic exited state decay rate, $2\kappa/2\pi =   9.5\pm0.1$ MHz is cavity full width at half maximum (FWHM) and $N_c$ is an effective number of atoms coupled to the cavity.   
Using an annular piezoelectric transducer (PZT) attached to one cavity mirror, the length of the cavity and hence its resonance frequency ($\omega_c$) can be tuned. On scanning $\omega_c$ around the atomic transition, with frequency $\omega_a$ (3-4$'$ transition of D2 line~\cite{steck_rubidium_2013}, which is also the cooling transition for the MOT), two peaks are observed in cavity emission. The cavity emission is monitored using a Photo Multiplier Tube (PMT)  and a CCD camera. One peak is red detuned side, and another peak is formed blue detuned to the atomic transition at $\omega_a$ as shown in Fig.~\ref{atnogrid}(a). Both these peaks show TEM$_{00}$ spatial mode structure imaged in the camera, i.e., the cavity output has Gaussian intensity profile. Other spatial modes also show similar two peaked behaviour.
The peak heights and frequencies of both the peaks depend on $N_c$ and power in the MOT beams. The red detuned peak is only visible above a critical number of atoms ($N_c \sim 20 \times 10^3$) as seen Fig.~\ref{atnogrid}(a) and grows rapidly to dominate the blue detuned peak for atom number $N_c \sim 33\times 10^3$. In addition, from Fig.~\ref{atnogrid}(a) the peaks are seen to broaden as $N_c$ increases.     

Fig.~\ref{atnogrid}(b) shows change in peak separation when $N_c$ is changed, as the cavity length is changed to scan across the atomic resonance frequency. Blue dots in Fig.~\ref{atnogrid}(b) show corresponding vacuum Rabi splitting (VRS) calculated for the same number of atoms and cavity parameters, when the cavity is on atomic resonance, and external weak probe light scans the atom-cavity system. From this close agreement between the measured frequency split and the VRS calculation we can see that the observed split depends on $g_0\sqrt{N_c}$, similar to VRS and hence we conclude that the observed splitting and the atom-cavity collective strong coupling are very closely related. However, this is the case only beyond a certain atom number in the cavity mode. Below a critical value of $N_c$, the red-detuned peak disappears while the blue detuned peak persists, as seen in Fig.~\ref{atnogrid}(c). Here a clear threshold is seen in the measured heights of the red detuned peak, while the blue detuned peak shows no threshold over the measurement range. It is also evident from Fig.~\ref{atnogrid}(b) and Fig.~\ref{atnogrid}(c) that the agreement with the VRS calculation starts degrading at the larger values of $N_c$, which coincides with the tendency towards saturation of the red detuned peak height. 
\onecolumngrid

\begin{figure}[H]
\centering
\includegraphics[width=14.0cm]{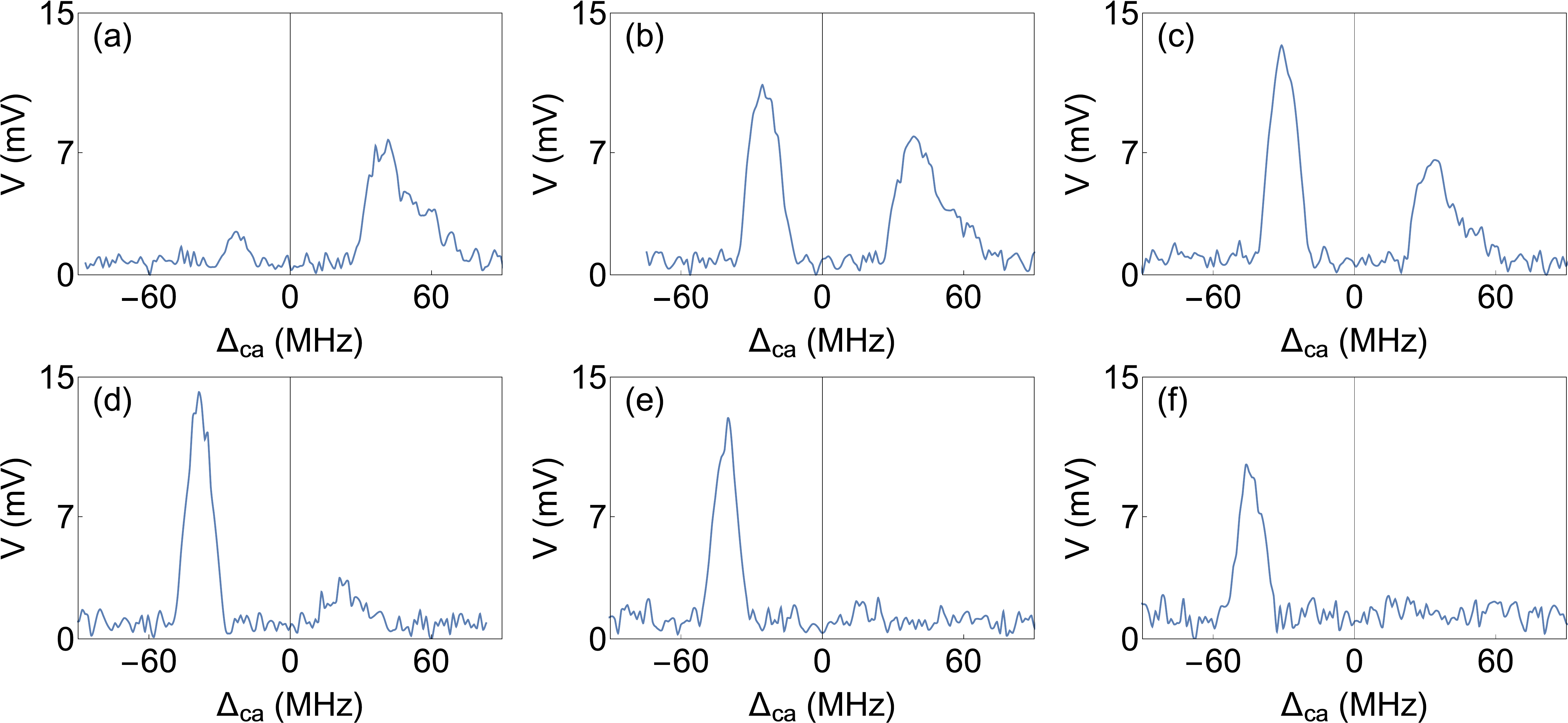}
\caption{Change in emission from the cavity when total intensity of MOT laser is varied. Intensities and atoms coupled ($N_c$) are (a) 22 mW/cm$^2$  and $(35 \pm 1) \times 10^3$, (b) 30 mW/cm$^2$ and $(33 \pm 1) \times 10^3$, (c) 36 mW/cm$^2$ and $(29 \pm 1) \times 10^3$, (d) 45 mW/cm$^2$ and $(24 \pm 1) \times 10^3$, (e) 60 mW/cm$^2$ and $(21 \pm 1) \times 10^3$, (f) 75 mW/cm$^2$ and $(23 \pm 1) \times 10^3$ respectively.}
\label{motpowgrid}
\end{figure}
\twocolumngrid
Fig.~\ref{motpowgrid} shows how the peaks change when the power of the MOT laser is increased. For very low power the red detuned peak vanishes and when the power in MOT laser is high the blue detuned peak vanishes. Increasing the power increases total atom number but also increases the spatial size of the atomic cloud in MOT. The increase in size is attributed to an increase in temperature of the MOT atoms~\cite{kowalski_magneto-optical_2010, petrich_behavior_1994, townsend_phase-space_1995}. The increase in size lowers $N_c$, and as fewer atoms overlap the cavity mode, the rate of fluorescence into the cavity decreases. Here, along with the change in $N_c$, we can expect to see a competition between the light emitted in the cavity mode by the atoms with the stimulated emission of the excited atoms by the MOT beams. This is because we go well above the saturation intensity of the atoms, even at the detuned frequencies of the MOT beams, for the case in Fig.~\ref{motpowgrid}. In this case, the spontaneous emission that the blue detuned peak represents competes poorly against stimulated emission into the MOT beams, compared with the red detuned peak. This provides experimental evidence of crucial differences between the blue and red detuned emission peaks observed.

%
\onecolumngrid

\begin{figure}[H]
\centering
\includegraphics[width=15cm]{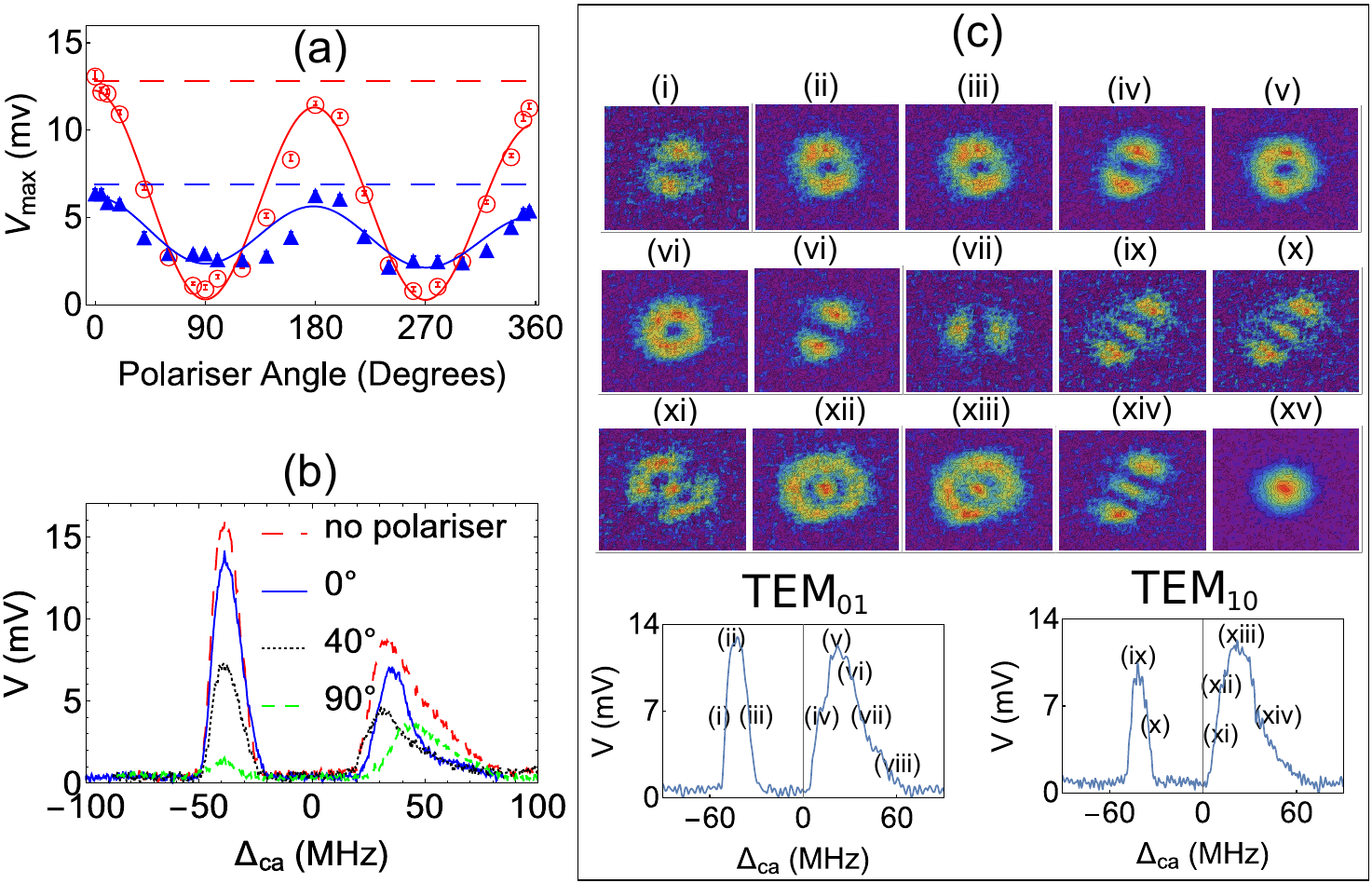}
\caption{Mode purity measurements. (a) Polarisation diagnosis of the light coming out of the cavity. Blue empty squares denote the height of the blue detuned peak, and red dots denote the height of the red detuned peak. Red dashed and blue dashed are peak values when we do not have a polariser in the cavity output path multiplied by transmission coefficient of the polariser. The smooth curves are obtained by fitting cosine function multiplied by a linear function, $\left(a+b\cos{[\frac{\pi}{180}x]}\right)\left(1-c x\right)$. The linear function is used to correct for the systematic effect of atom number decrease during long measurements. (b) Cavity output profile for different polariser angles. Here, $N_c = (26 \pm 1) \times 10^3$ for the above graphs. (c) Modes of the cavity as seen on a CCD camera at various detunings of the cavity. Lower case Roman numerals (i)-(viii) are for TEM$_{01}$ mode and (ix)-(xiv) are for TEM$_{10}$ mode. (xv) shows an instance of TEM$_{00}$ mode. Here, for TEM$_{00}$ mode $N_c = (33 \pm 1) \times 10^3$, it will be lower for higher order modes.  The power of MOT cooling laser is  36 mW/cm$^2$ for all the cases.}
\label{pol}
\end{figure}
\twocolumngrid
\section{Signatures of lasing}
The threshold behaviour seen only in the red detuned peak in Fig.~\ref{atnogrid}(c) is suggestive of lasing in this driven atomic system coupled to the cavity mode. Since during lasing one mode experiences large gain compared to others as stimulated emission ensures that the dominant mode wins, so to check for lasing by the atom-cavity system, we check for purity of the state of light that comes out of the cavity. The first check is for the polarisation state of the cavity emitted light using a polariser (Thorlabs LPNIRE100-B) with variable polarisation angle in between the cavity mirror and the PMT. Fig.~\ref{pol}(a) shows the change in peak height of both red and blue detuned peaks as the polarization angle is changed. The  polarisation of the red detuned peak is considerably purer than that of the blue detuned peak. From the fits to data, the visibility for the red detuned peak is around 95 \% and for the blue detuned peak is around 43 \%. The small change in amplitude with polariser angle is due to the slight decrease in atom number as the measurement progressed. Also from Fig.~\ref{pol}(b), it can be seen that the shape of the blue detuned peak depends on polariser angle. The peak and the wings of the blue detuned peak seem to be oppositely polarised. This shows that the light out of the cavity for the red peak is strongly polarised and indicative of a single cavity mode present in the cavity. On the other hand, the blue detuned peak has multiple modes in coexistence, which is indicative of no mode competition and therefore is detected as a mixture of several simultaneously existing modes. The consistent explanation for this is that the light out of the cavity for the red peak is lasing. All the data mentioned above was for TEM$_{00}$ spatial mode as there are no nearby modes within the cavity linewidth which will compete with this mode. This is verified by imaging the spatial mode emitted by the cavity. However, for higher spatial modes there is competition between the cylindrical and rectangular modes which are nearly degenerate as can be seen explicitly from Fig.~\ref{pol}(c). Here again for the red detuned peak a single dominant spatial mode is observed while the blue detuned cavity emission peak shows multiple spatial modes of similar intensities. The purity of polarisation and the spatial mode is strong evidence for lasing in the case of the red detuned peak. The sharp threshold for the red detuned peak and the rapid increase in peak height as the atom number is increased is characteristic of lasing behaviour arising due to non-linear response with increase in an gain~\cite{silfvast_laser_2004}.  

\begin{figure}[H]
\centering
\includegraphics[width=15cm]{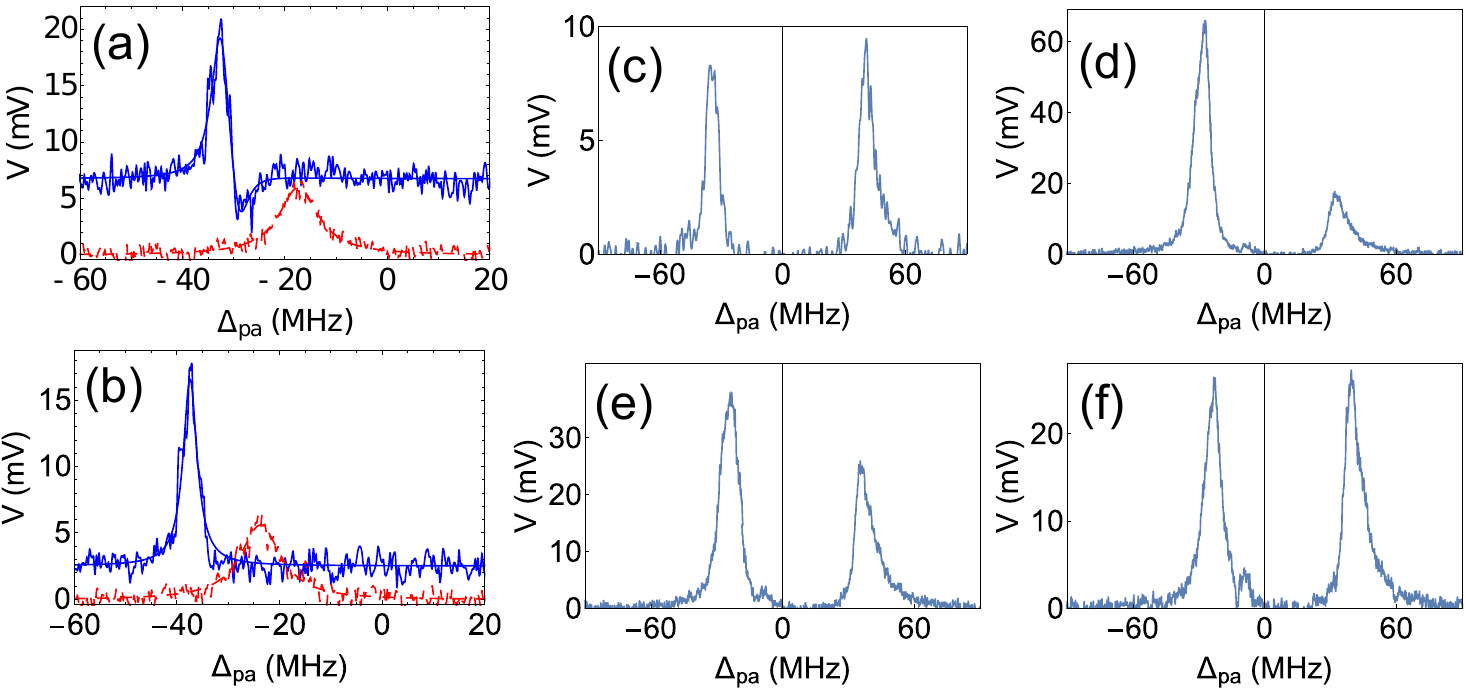}\caption{Probe transmission spectra when atoms are present inside the cavity. For (a) and (b), the blue trace shows cavity probe transmission when driven atoms are present in the cavity, and the broken red trace is probe transmission curve without atoms.(a) This figure shows a gain and Fano-type structure in the presence of atoms. The cavity length is such that we are near the top of the red detuned peak of the last graph in Fig.~\ref{atnogrid}(a). We can see a gain in the probe at around 33 MHz. The blue curve in the background is fit to the sum of Lorentzian and inverted Lorentzian both of FWHM 4.5 MHz, and the red dashed curve for the empty cavity is a Lorentzian with FWHM of 9.5 MHz. (b) This figure shows line narrowing and gain without the Fano structure when the cavity is slightly red of the red detuned peak of the last graph in Fig.~\ref{atnogrid}(a). A Lorentzian fit to the blue trace gives FWHM of 3.3 MHz and is shown as the blue curve. Here, for (a) and (b) $N_c = (33 \pm 1) \times 10^3$.
(c) Without MOT laser and cavity freq same as atomic.
(d) With MOT laser and cavity freq. same as atomic freq.
(e) With MOT laser and cavity freq. towards blue of atomic freq.
(f) With MOT laser and cavity freq. more towards blue of atomic freq to make the heights of the peak same. For (c)-(f), $N_c = (25 \pm 1) \times 10^3 $. The power in MOT laser for all the measurements in this figure is 36 mW/cm$^2$.}
\label{fano}
\end{figure}

To measure the frequency at which the gain occurs, we pass a probe beam through the cavity and monitor the transmission. Fig.~\ref{fano}(a) shows the transmitted intensity when the frequency of the probe beam is scanned while keeping the cavity detuning fixed on the red detuned peak of the last graph in Fig.~\ref{atnogrid}(a). The blue trace represents the output with MOT atoms, and the red trace is the output when there are no atoms in MOT (the magnetic field is switched off). As is evident from the differences in the transmission in Fig.~\ref{fano}(a), the probe beam experiences gain which is more than the total losses when interacting with atoms  inside the cavity. The shift between blue and the red curve is due to lasing without any probe laser as we make sure that the cavity detuning is fixed on the red peak of the last graph in Fig.~\ref{atnogrid}(a). In addition to gain, the blue trace also shows a Fano-like structure indicating interference between two or multiple probability amplitudes for emission~\cite{miroshnichenko_fano_2010} which suppresses the lasing action. Fig.~\ref{fano}(b) shows line narrowing even in the absence of Fano structure slightly away from the red peak of the last graph in Fig.~\ref{atnogrid}(a). A similar experiment on the blue detuned peak attenuates and shows a tendency to broaden the transmitted light. 
\onecolumngrid

\twocolumngrid
%
\section{Gain mechanism}

The phenomenon of gain in probe beam when two-level atoms are strongly driven by another drive laser is well known. B. R. Mollow was first to predict such gain mechanism~\cite{mollow_stimulated_1972} using semi-classical arguments and phenomenologically including atomic decay rates. Haroche \textit{et.al}~\cite{haroche_theory_1972} also did similar calculations during the same year.  This was later observed in an experiment with sodium beam by Wu \textit{et.al}~\cite{wu_observation_1977}. Grynberg \textit{et.al}~\cite{grynberg_central_1993} later calculated the cross-section for such gain using microscopic perturbation theory. They showed that in the presence of a strong laser the atomic states get dressed, and the gain arises because of multi-photon processes between the dressed states. In the case where the driving laser is red detuned, this type of gain occurs only when the frequency of probe is to the red side of the drive laser frequency. For blue detuned peak such multi-photon processes lead to loss due to absorption. The mechanism for the gain here is similar to the work discussed above, with the additional consideration that the cavity-atom collective strong coupling regime introduces certain differences discussed below.

In the regime defined by collective strong coupling, weak probe transmission through the cavity is split into two peaks when atoms are present in the cavity. This effect is called VRS and is a well-known phenomenon~\cite{thompson_observation_1992,agarwal_spectroscopy_1998,hernandez_vacuum_2007}. Fig.~\ref{fano}(c) is a measurement of such a split for which the MOT lasers were switched off, and all the atoms were in the ground state~\cite{ray_temperature_2013}. Here, if the cavity resonance frequency is matched with the atomic frequency we see a symmetric splitting on the red and the blue side of the atomic resonance. However, if the cavity resonance is blue (red) detuned with respect to the atomic frequency the blue (red) detuned VRS peak is bigger than red (blue) detuned one~\cite{agarwal_spectroscopy_1998} and blue (red) peak shifts away (towards) from atomic frequency. However, when the drive laser with a frequency different from atomic transition is switched on, symmetry is broken because of unequal gain and loss. Fig.~\ref{fano}(d,e,f) shows such a scenario. For Fig.~\ref{fano}(d), when the MOT drive lasers are present, the frequency spectrum remains unchanged, but the heights are completely different, indicating an unequal total loss. The drive (MOT cooling laser) is to the red side of atomic frequency. For Fig.~\ref{fano}(e) even though the cavity resonance is set to the blue of the atomic resonance, the red detuned peak is dominant. In the absence of drive laser, the height of blue detuned peak would have increased, and the height red detuned peak would have diminished. However, due to Mollow gain in the the red detuned probe and Mollow loss in the blue detuned probe in the presence of the red detuned drive we see the opposite in Fig.~\ref{fano}(e) with respect to what is expected in the case when the atoms are not optically driven. Equal height for the two peaks is obtained when the cavity resonance is 8MHz detuned towards blue of atomic frequency as shown in Fig.~\ref{fano}(f). In addition to this there also seems to be gain around -13 MHz which is the drive frequency. This is because of coherent energy exchange between drive and cavity. The observations mentioned above are explained using a semi-classical theory in the discussion section.             

In addition to gain in the probe, we also observe lasing without any seeding (no probe light). This is because of the coupling of light into the cavity by the fluorescing atoms. B. R. Mollow was first to calculate the fluorescence spectrum of a two-level atom driven by a strong monochromatic laser~\cite{mollow_power_1969}. Such a spectrum can be computed from fluctuations of atomic coherence~\cite{steck_quantum_2011} and is shown in the Fig.~\ref{mollow}. There is some light at $\sim$ 30 MHz red of the atomic transition. This resonance fluorescence provides the first photon which is amplified by the driven atoms. 
 
%
%
\section{Discussion}
To explain and interpret why the gain and lasing in the red-detuned peak from the driven-atom, cavity coupled system occurs we discuss the results of a model developed to understand the system. The model contains all the essential elements required to understand the lasing observed. We consider 2-level atoms coupled to one of the modes of a Fabry-P\'erot cavity, which is probed by a monochromatic laser of frequency $\omega_p$ and the atoms are driven by a single drive laser with frequency $\omega_d$. The probe and drive light fields are assumed to be classical.  

\begin{figure}[!t]
	\centering
	\includegraphics[width=8.5cm]{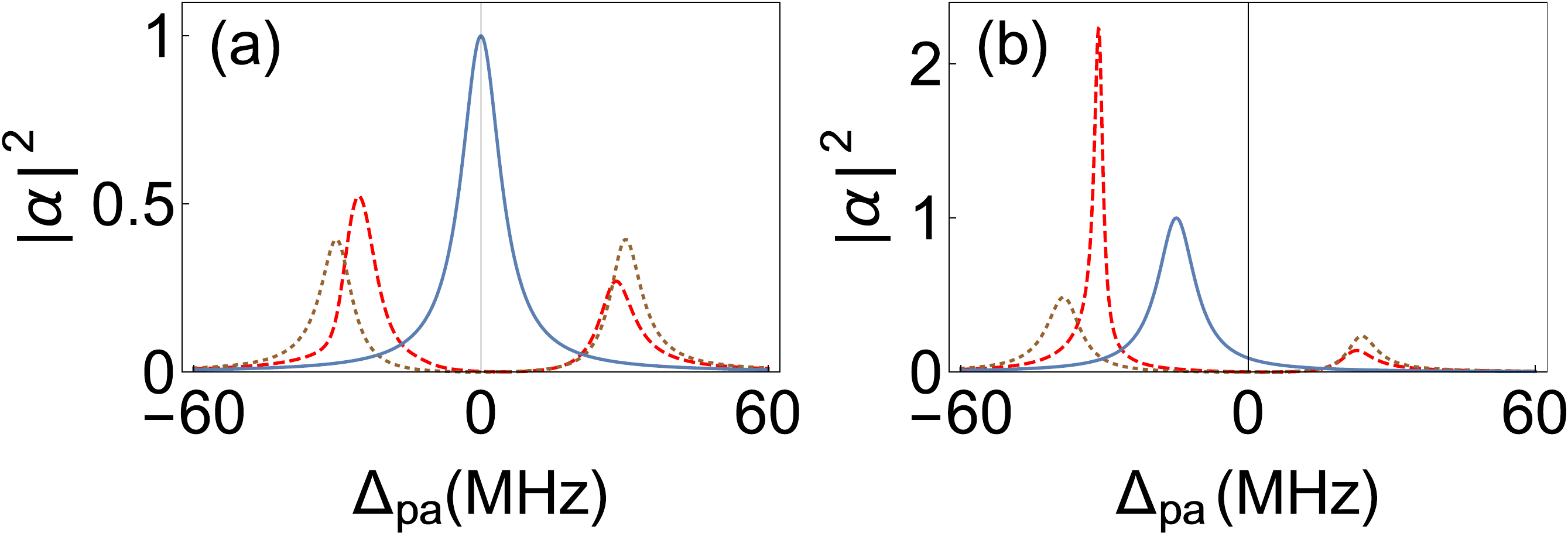}
	\caption{Probe transmission through the cavity as a function of detuning from the atomic transition. Blue is empty cavity transmission, the dotted brown curve is the VRS and red dashed curve shows probe transmission when atoms are driven by classical light. The two graphs are for different detuning of the cavity from atomic transition, (a) $\Delta_{ca} = 0$ and (b) $\Delta_{ca} =-15$ MHz. Other parameters are strength of drive field, $\Omega/(2\pi) = 6$ MHz, strength of probe field $\eta/(2\pi) = 4.75$ MHz, atom-cavity collective strong coupling, $g_t/(2\pi) = 30$ MHz, $\Gamma/(2\pi) = 6$ MHz, cavity FWHM, $2\kappa/(2\pi) = 9.5$ MHz, and $\Delta_{pa} =-	13$ MHz. $\alpha$ is field amplitude inside the cavity. The photon number is equal to $|\alpha|^2$.}
	\label{gainsim}
\end{figure}

The average photon number in the cavity is calculated using a semi-classical theory of atom-light interaction. Fig.~\ref{gainsim} shows cavity transmission spectrum as a function of the frequency of probe laser using Eqn.~\ref{alphafull}. From Fig.~\ref{gainsim} it is evident that for red detuning the probe laser experiences gain and for blue detuning experiences loss. This is completely consistent with the experimental observations in Fig.~\ref{fano}. When the gain of probe laser from atoms exceed the loss from the atoms-cavity system, we see lasing as shown in Fig.~\ref{gainsim}(b) providing an explanation for the lasing of the red-detuned peak. In addition, the parameters in our calculations are such that $4|\Omega|^4 < |\Delta_{da}|^3\Gamma$ and $4|\Omega|^2 < |\Delta_{da}|^2$, where $2\Omega$ is Rabi frequency of the drive laser, $\Delta_{da}$ is detuning of drive laser from the atomic transition and $\Gamma$ is decay rate of the atomic excited state due to spontaneous emission. This is opposite to the conditions necessary for gain in probe beam according to~\cite{grynberg_central_1993} and~\cite{mollow_stimulated_1972} respectively, where the system was required to be driven hard. However, here the presence of cavity enhances the interaction, and we see gain even without the very demanding requirements on parameters in the earlier experiments. 
On changing the detuning of the cavity with respect to the atomic transition, the collective eigenstate (VRS) also shifts in frequency. When the cavity is blue detuned with respect to the atomic transition, the red VRS peak shifts towards the atomic transition and the blue peak shifts away from the atomic transition. This makes it possible for the red VRS peak to be in resonance with the MOT cooling laser which is 13 MHz red detuned from the atomic transition when the detuned cavity resonance is to the blue side. As can be inferred from Fig.~\ref{mollow} the atomic fluorescence is dominant at MOT/drive laser frequency. Hence, the combined system of atoms and cavity becomes resonant for most of the fluorescent light, when the cavity is blue detuned with respect to the atomic transition. This explains the observation of the blue peak in the experiment as shown in the Figs.~\ref{atnogrid}(a), \ref{motpowgrid} and \ref{pol}(b). This blue peak is unpolarised as the fluorescence can be of any polarisation.

Comparing the numbers in the above calculations and the experiment, the total intensity of all the MOT beams and dipole moment for isotropic beam case gives $\Omega/(2\pi) \approx 6$ MHz as an estimate. 
However, theoretical treatment is for single drive beam whereas there are six beams, all with different polarisation and non-zero relative phase difference between them. The calculation therefore is a lower estimate of the light-atom interaction, since sub-Doppler cooling mechanisms will try to localise atoms towards intensity maxima formed by the six MOT beams increasing effective value of $\Omega$~\cite{townsend_phase-space_1995}. In addition, for the analysis we have assumed the cavity field to be very weak, such that the atomic coherence, $\rho$ is only first order in cavity field, $\alpha$. There is a possibility of gain to be higher order in $\alpha$ which has not been considered here. Hence exact quantitative comparison is not possible with the present model, but qualitatively we do not expect the physics behind the observations to change.   

%
\section{Conclusion}
In conclusion, when a MOT of actively driven atoms is collectively strongly coupled to a Fabry-P\'erot cavity, lasing is seen in the red detuned emitted light from the atom-cavity system. The lasing for a stabilized cavity and MOT atom number is continuous in nature. 
This driving field is the same as the MOT light which cools and traps the atoms and results in continuous operation of the laser from the coupled atom-cavity system. The lasing is seen for the red detuned peak because symmetry is broken in favour of the red detuned peak due to the MOT light, which also confines the atoms effectively and couples it to the cavity mode. In principle the lasing could be achieved on the blue detuned peak if the driving frequency and the trap mechanism were not connected and the driving light was blue detuned with respect to the atomic transition. The observed phenomena are therefore generic to the driven atom-cavity system and have potential to be of interest for basic studies and for applications. To analyse the basic processes of this phenomena a semiclassical theory of light-matter interaction is presented in the discussion section and appendix, where the role of Mollow gain and dispersion due to the collective strong coupling of atoms to the cavity and the change in the dynamics of the system is explained. Here, we took the decay of the atomic excited state phenomenologically similar to B. R. Mollow~\cite{mollow_stimulated_1972}. Theoretically, it would be interesting to know how the laser properties change if a full quantum and multimode description including the vacuum modes~\cite{grynberg_central_1993} is used. The vacuum modes and the drive field can be then integrated out similar to the approach of open quantum optics~\cite{scully_quantum_1997} giving an effective non-linear interaction between the photons in the cavity and the atoms. Future work can be taken up to observe the properties of this type of non-linear laser, such as photon
statistics, phase diffusion and squeezing, as predicted by single mode theories of Zakrzewski \textit{et.al}~\cite{zakrzewski_theory_1991,zakrzewski_theory_1991-1,zakrzewski_theory_1991-2} and Agarwal~\cite{agarwal_theory_1990,agarwal_dressed-state_1990}                 

\onecolumngrid 
\appendix*
\section{}
\subsection{Details of calculations}\label{calculations}

The Hamiltonian of a system of driven $N$ two-level atoms interacting with a cavity driven by a probe laser can be written as,

\begin{align}
\hspace{-0.3 cm}\hat{H}&=\hbar \omega_c \ \hat{a}^{\dagger}\hat{a} +\hbar\left\{\eta^* \hat{a} e^{-i\omega_p t}+\eta \hat{a}^{\dagger} e^{i\omega_p t} \right\} + \sum^{N}_{j}\hbar\left\{\frac{\omega_a}{2}\hat{\sigma}^{z}_j   + g_j \left(\hat{a}^{\dagger} \hat{\sigma}^{-}_j + \hat{a} \hat{\sigma}^{+}_j \right)\right\} +\sum^{N}_{j}\hbar\left\{\Omega^{*} \hat{\sigma}^{-}_j e^{i\omega_d t}+ \Omega \hat{\sigma}^{+}_j e^{-i\omega_d t} \right\}.
\label{ham}
\end{align}
$\hat{a}$ and $\hat{a}^{\dagger}$ are photon annihilation and creation operators for the cavity field with commutation relation $[\hat{a},\hat{a}^{\dagger}] =1$. ${\sigma}^{-}_j$, ${\sigma}^{+}_j$, ${\sigma}^{z}_j$ are usual spin-$\frac{1}{2}$ Pauli operators for the $j^{th}$ atom with commutation relations $[{\sigma}^{+}_j,{\sigma}^{-}_j] ={\sigma}^{z}_j$ and $[{\sigma}^{z}_j,{\sigma}^{\pm}_j] =\pm{\sigma}^{\pm}_j$. $g_j = g_0 f(x_j,y_j)$ is coupling of $j^{th}$ atom with the cavity mode with $g_0=-\mu\sqrt{\omega_c /(2 \hbar \epsilon_0 V)}$ maximum atom-cavity coupling, $f(x,y) = e^{-(x^2+y^2)/w_0^2}$ is mode function of the cavity field, $V = \pi w_0^2L_c/4$ is the mode volume of the cavity, $L_c$ is cavity length, $\mu$ is the transition dipole matrix element between excited and ground state, $2|\Omega|=\frac{\mu}{\hbar}\sqrt{\frac{2I_p}{c\epsilon_0}}$ is Rabi frequency for drive beam, $I_p$ is intensity of the drive beam, $\eta$ is strength of classical cavity probe. $\omega_c$, $\omega_a$, $\omega_d$, $\omega_p$ are frequencies of the cavity (resonance frequency of cavity when it is empty), atom, driving laser and cavity probe respectively and $^{*}$ denotes complex conjugate. 

\subsubsection{Rate equations}
The evolution equation for the expectation value for an operator $\hat{X}$ can be evaluated using the Heisenberg equation, $\frac{d\langle\hat{X}\rangle}{dt} = \frac{i}{\hbar} \langle [\hat{H},\hat{X}]\rangle$.
In addition to this unitary evolution, we introduce non-unitary decay rates $\kappa$ and $\Gamma$ phenomenologically for cavity field and atoms respectively. This approach is similar to the one taken by B. R. Mollow~\cite{mollow_stimulated_1972}. For the calculation here, the cavity field is assumed to be classical and is denoted by a coherent state $\ket{\alpha}$.
Using the Heisenberg equation and decay rates mentioned above for system variables give, 
\begin{subequations}\label{timd}
\begin{align}
&\frac{d\alpha(t)}{dt} =  -\kappa \alpha(t) - i\omega_c \alpha(t)  - i\sum^{N}_{j}g_j \rho_j(t) - \eta e^{-i\omega_p t} \\
&\frac{d\rho_j(t)}{dt} = -\left\{\frac{\Gamma}{2} + i \omega_a \right\} \rho_j(t) + i( g_j \alpha(t)+\Omega e^{-i\omega_d t}) (2\rho^e_{j}(t) - 1)   \\
&\frac{d\rho_{e,j}(t)}{dt} =  -\Gamma \rho_{e,j}(t)+ i\left\{( g_j \alpha^{*}(t)+\Omega^{*} e^{	i\omega_d t}) \rho_{j}(t) - g_j \alpha(t)+\Omega e^{-i\omega_d t}) \rho_{j}^{*}(t) \right\}
\end{align}
\end{subequations}
$\rho_j = \expect{\sigma^-_j}$ is coherence of the $j^{th}$ atom, $\alpha = \expect{\hat{a}}$ is field amplitude inside the cavity and $\rho_{e,j} = \expect{\sigma_{e,j}}$ is excited state population.

As multiple frequencies are involved, we transform the time-dependent equations~(\ref{timd}) into Fourier space. 
For the cavity field and the atomic coherences, the Fourier transforms are,
$\alpha(t) = \int_{-\infty}^{\infty}\tilde{\alpha}(\omega) e^{-i\omega t} d\omega$ and 
$\rho_j(t) = \int_{-\infty}^{\infty}\tilde{\rho}_j(\omega) e^{-i\omega t}d\omega$ respectively. 
And as $\rho_{e,j}(t)$ is always real, its Fourier transform takes the form, $\rho_{e,j}(t)=\int_{-\infty}^{\infty}\frac{1}{2}(\tilde{\rho}_{e,j}(\omega)e^{-i\omega t}+\tilde{\rho}^*_{e,j}(\omega)e^{i\omega t})d\omega$ with the relation $\tilde{\rho}^*_{e,j}(-\omega) = \tilde{\rho} _{e,j}(\omega)$. This relation can be verified by replacing $\omega$ with $-\omega$ in Fourier relation of $\rho_{e,j}(t)$.
The inverse transforms are, $\tilde{\alpha}(\omega)  = \int_{-\infty}^{\infty}\alpha(t)e^{-i\omega t}dt$, 
$\tilde{\rho}_j(\omega)  = \int_{-\infty}^{\infty} \rho_j(t)e^{i\omega t}dt$.
$\tilde{\rho}_{e,j}(\omega)=\int_{-\infty}^{\infty}\rho_{e,j}(t) e^{i\omega t}dt$ and $\tilde{\rho}^*_{e,j}(\omega)=\int_{-\infty}^{\infty}\rho_{e,j}(t) e^{-i\omega t}dt$.\\
This gives a set of equations,
\begin{subequations}\label{fd}
\begin{align}
&-i\omega \tilde{\alpha}(\omega) =  -\kappa\tilde{\alpha}(\omega) - i\omega_c \tilde{\alpha}(\omega)  - i\sum^{N}_{j}g_j \tilde{\rho}_j(\omega) -\eta \delta(\omega-\omega_p) \\
&-i\omega\tilde{\rho}_j(\omega) = -\left\{\frac{\Gamma}{2} + i \omega_a \right\} \tilde{\rho}_j(\omega) + ig_j \left\{2 (\tilde{\alpha}\odot\tilde{\rho}^*_{e,j})(\omega)_{-} - \tilde{\alpha}(\omega) \right\} + i\Omega \left\{2 \tilde{\rho}^*_{e,j}(\omega_d-\omega) - \delta(\omega - \omega_d) \right\}    \\
&i\omega\tilde{\rho}^*_{e,j}(\omega)=  -\Gamma \tilde{\rho}^*_{e,j}(\omega)  + ig_j\left\{(\tilde{\alpha}^{*}\odot\tilde{\rho}_j)(\omega)_{-} - (\tilde{\alpha}\odot\tilde{\rho}^{*}_j)(\omega)_{+} \right\} + i\left\{\Omega^*\tilde{\rho}_j(\omega_d - \omega) -\Omega\tilde{\rho}^*_j(\omega_d + \omega) \right\} 
\end{align}
\end{subequations}

Here,
eqn~(\ref{timd}a) and eqn~(\ref{timd}b) are multiplied by $e^{i\omega t}$ and  eqn~(\ref{timd}c) by $e^{-i\omega t}$ and then integration  $\int_{-\infty}^{\infty} dt$ is done. $(\tilde{\alpha}\odot\tilde{\rho}^{*}_j)(\omega)_{\pm} = \int\tilde{\alpha}(\omega_1)\tilde{\rho}^{*}_j(\omega_1\pm\omega)d\omega_1$ are convolution functions where $\odot$ denotes convolution operation. $\delta$ is Dirac delta function.

To understand the physics we discuss the various interactions sequentially. First, we neglect the classical driving of the atoms, i.e. $\Omega = 0$ and assume that the cavity field (probe laser) is so weak that the atoms remain mostly in ground state, i.e $\tilde{\rho}_{e,j}(\omega) = \tilde{\rho}^*_{e,j}(\omega) \approx 0$. 
 For such a case the cavity field is,
\begin{align}
&\tilde{\alpha}(\omega_d)=  \frac{-\eta\left\{\frac{\Gamma}{2} - i \Delta_{pa} \right\}}{\left\{\kappa- i\Delta_{pc} \right\} \left\{\frac{\Gamma}{2} - i \Delta_{pa} \right\}  + g_t^2}
 \label{cavonly}
 \end{align}
$g_t =\sqrt{\sum^{N}_{j}g_j^2} $ is total g for all the atoms and is equal to $g_0 \sqrt{N_c}$, where $g_0$ is the single atom-cavity mode coupling and $N_c$ is an average number of atoms coupled to the cavity and can be calculated using an overlap of the cavity mode and the atomic density distribution. $\Delta_{pc} = \omega_p - \omega_c$ and $\Delta_{pa} = \omega_p - \omega_a$ are detunings of the cavity probe laser from the cavity and the atomic frequencies respectively. Eqn.~(\ref{cavonly}) gives the usual splitting in the cavity peak (VRS/normal mode splitting). The splitting between two peaks is equal to $2g_t = 2g_0 \sqrt{N_c}$. 

Now, in the absence of a cavity all the atoms behave in same way. This is because all are driven equally by the classical drive field. 
The atomic variables in this case are,
\begin{align}
&\tilde{\rho}(\omega_d) = \frac{-2i\Omega(2 i \Delta_{da}+ \Gamma)}{\Gamma ^2+4 \Delta_{da}^2+8 |\Omega|^2} \ \ \ \text{and} \ \ \ \tilde{\rho}_e(0) =\frac{4 |\Omega|^2}{\Gamma ^2+4 \Delta_{da}^2+8 |\Omega|^2}
\label{cohnocav}
\end{align}   
$\Delta_{da} = \omega_d - \omega_a$ is detuning of drive laser from the atomic transition. The $j$ subscript is removed as the atom-field interaction strength is same for all the atoms. The atomic coherence oscillates only at frequency $\omega_d$ and all other frequency components are zero. 

\subsubsection{Perturbative calculation}
For our experiment, $\Omega^2 \gg g_0^2 |\alpha|^2$ and hence the effect of the cavity on the atoms is very small and adds only small perturbations to values in (\ref{cohnocav}). The perturbed atomic variables can be written as,
$\tilde{\rho}_j(\omega) = \tilde{\rho}(\omega_d)\delta(\omega-\omega_d) + \tilde{\epsilon}_j(\omega)$, $\tilde{\rho}_{e,j}(\omega) = \tilde{\rho}_e(0)\delta(\omega) + \tilde{\epsilon}_{e,j}(\omega)$and $\tilde{\rho}^*_{e,j}(\omega) = \tilde{\rho}_e(0)\delta(\omega) + \tilde{\epsilon}^*_{e,j}(\omega)$. Here, $\epsilon$ is perturbation to density matrix elements of atoms due to interaction with the cavity.   
Keeping only the unperturbed part in the convolution functions gives a set of linear equations, 
\begin{subequations}\label{fd1}
\begin{align}
&-i\omega \tilde{\alpha}(\omega) =  -\kappa\tilde{\alpha}(\omega) - i\omega_c \tilde{\alpha}(\omega)  - i\sum^{N}_{j}g_j \tilde{\rho}_j(\omega) -\eta \delta(\omega-\omega_p) \\
&-i\omega\tilde{\rho}_j(\omega) = -\left\{\frac{\Gamma}{2} + i \omega_a \right\} \tilde{\rho}_j(\omega) + ig_j \tilde{\alpha}(\omega) \left\{2 \tilde{\rho}_{e}(0) - 1 \right\} + i\Omega \left\{2 \tilde{\rho}^*_{e,j}(\omega_d-\omega) - \delta(\omega - \omega_d) \right\}    \\
&i\omega\tilde{\rho}^*_{e,j}(\omega)=  -\Gamma \tilde{\rho}^*_{e,j}(\omega) + ig_j\left\{\tilde{\alpha}^{*}(\omega_d +\omega)\tilde{\rho}(\omega_d) - \tilde{\alpha}(\omega_d -\omega)\tilde{\rho}^{*}(\omega_d) \right\} + i\left\{\Omega^*\tilde{\rho}_j(\omega_d - \omega) -\Omega\tilde{\rho}^*_j(\omega_d + \omega) \right\} 
\end{align}
\end{subequations}

In zero$^{th}$ order of the atomic terms, the cavity field gets an additional term,
\begin{align}
&i\sum^{N}_{j}g_j \tilde{\rho}_j(\omega) = i\sum^{N}_{j}g_j \left\{\tilde{\rho}(\omega_d)\delta(\omega-\omega_d) + \tilde{\epsilon}_j(\omega)\right\} \approx i\sum^{N}_{j}g_j \tilde{\rho}(\omega_d)\delta(\omega-\omega_d) 
\label{zerot}
\end{align}
This requires $\omega_p = \omega_d$ as seen in eqn~(\ref{zerot}) and results in an elastic exchange of energy from drive field to cavity field via atoms and the cavity will gain photons even if it is not driven by an external field.    

For other frequencies the perturbation term $\tilde{\epsilon}_j(\omega)$ is important, hence we proceed to derive it using Eqn~(\ref{fd1}). 
Eqn~(\ref{fd1}b) can be rewritten as,
\begin{align}
 & \hspace{-0.3 cm} -\left\{\frac{\Gamma}{2} + i \Delta_a(\omega) \right\}\tilde{\epsilon}_j(\omega) + ig_j\tilde{\alpha}(\omega) \left\{2 \tilde{\rho}_e(0)  - 1 \right\} + i\Omega \left\{2  \tilde{\epsilon}^*_{e,j}(\omega_d-\omega) \right\} \nonumber \\
& \hspace{-0.3 cm} = \left( \left\{\frac{\Gamma}{2} + i \Delta_a(\omega) \right\} \tilde{\rho}(\omega_d)- i\Omega \left\{2 \tilde{\rho}_e(0)  - 1 \right\} \right) \delta(\omega - \omega_d) 
\end{align}
The right-hand side of above equation is zero for all $\omega$ because the term $\tilde{\rho}(\omega_d)$ was derived by putting the linear equation in right-hand side bracket with $\omega = \omega_d$ to zero in the absence of a cavity as can be inferred from Eqn~(\ref{fd}) by putting the cavity field terms to zero.
This gives,
 \begin{align}
&\hspace{-0.3 cm}-\left\{\frac{\Gamma}{2} + i \Delta_a(\omega) \right\}\tilde{\epsilon}_j(\omega) + ig_j\tilde{\alpha}(\omega) \left\{2 \tilde{\rho}_e(0)  -1 \right\} + i\Omega \left\{2  \tilde{\epsilon}^*_{e,j}(\omega_d-\omega) \right\} = 0
\label{sigma0}
\end{align}    
Using~(\ref{sigma0}) and $\tilde{\epsilon}^*_{e,j}(\omega_d-\omega) \approx 0$ gives first order term for  $\tilde{\epsilon}_j(\omega)$,
\begin{align}
&\tilde{\epsilon}_j(\omega_p)= -ig_j\frac{\tilde{\alpha}(\omega_p) \left\{-2 \tilde{\rho}_e(0)  +1 \right\}}{\left\{\frac{\Gamma}{2} - i \Delta_{pa} \right\}} = -ig_j\tilde{\alpha}(\omega_p)C_1(\omega_p)
\label{firstap}
\end{align}
This is similar to the value of $\tilde{\rho}(\omega)$ used to calculate Eqn~(\ref{cavonly}) but with an extra factor of $\left\{-2 \tilde{\rho}_e(0)  +1 \right\}$. This factor comes because of change in  population difference between excited and ground state and effectively reduces the coupling as the cavity field sees fewer atoms in the ground state. 

Now for excited state population if we do not neglect the term, $\tilde{\epsilon}^*_{e,j}(\omega_d-\omega)$,
\begin{align}
 &\hspace{-0.2 cm} (i\omega_d - i\omega +\Gamma)[\tilde{\rho}_e(0)\delta(\omega_d -\omega) +\tilde{\epsilon}^*_{e,j}(\omega_d -\omega)]\nonumber \\
&\hspace{-0.2 cm} =   ig_j\left\{\tilde{\alpha}^{*}(2\omega_d-\omega)\tilde{\rho}(\omega_d) - \tilde{\alpha}( \omega)\tilde{\rho}^{*}(\omega_d) \right\} +i\left\{\Omega^*\tilde{\rho}_j(\omega) -\Omega\tilde{\rho}^*_j(2\omega_d-\omega) \right\} 
\end{align}
After rearranging we get, 
\begin{align}
& \hspace{-0.3 cm} (i\omega_d - i\omega +\Gamma)\tilde{\epsilon}^*_{e,j}(\omega_d -\omega) - ig_j\left\{\tilde{\alpha}^{*}(2\omega_d-\omega)\tilde{\rho}(\omega_d) - \tilde{\alpha}( \omega)\tilde{\rho}^{*}(\omega_d) \right\} - i\left\{\Omega^*\tilde{\epsilon}_j(\omega) -\Omega\tilde{\epsilon}^*_j(2\omega_d-\omega) \right\} \nonumber \\
&\hspace{-0.3 cm}=\left[-(i\omega_d - i\omega +\Gamma)\tilde{\rho}_e(0)  + i\left\{\Omega^*\tilde{\rho}_j(\omega_d) -\Omega\tilde{\rho}^*_j(\omega_d) \right\}\right]\delta(\omega_d -\omega) 
\label{sigmae0}
\end{align}
Again the right hand side is zero for all $\omega$.
The complete set of equations to calculate $\tilde{\epsilon}_j(\omega_p)$ at the cavity probe frequency can be derived from~(\ref{sigma0}) and~(\ref{sigmae0}) and are given by,
\begin{subequations}\label{set1}
\begin{align}
&\hspace{-0.9 cm}-\left\{\frac{\Gamma}{2} + i \Delta_a(\omega_p) \right\}\tilde{\epsilon}_j(\omega_p) + ig_j\tilde{\alpha}(\omega_p) \left\{2 \tilde{\rho}_e(0)  -1 \right\}+  i\Omega \left\{2  \tilde{\epsilon}^*_{e,j}(\omega_d-\omega_p) \right\}=0\\
&\hspace{-0.9 cm}-[i(\omega_d-\omega_p) +\Gamma]\tilde{\epsilon}^*_{e,j}(\omega_d -\omega_p)+ ig_j\left\{\tilde{\alpha}^{*}(2\omega_d-\omega_p)\tilde{\rho}(\omega_d) - \tilde{\alpha}( \omega_p)\tilde{\rho}^{*}(\omega_d) \right\} + i\left\{\Omega^*\tilde{\epsilon}_j(\omega_p) -\Omega\tilde{\epsilon}^*_j(2\omega_d-\omega_p) \right\}=0\\
&\hspace{-0.9 cm}-\left\{\frac{\Gamma}{2} + i \Delta_a(2\omega_d-\omega_p) \right\}\tilde{\epsilon}_j(2\omega_d-\omega_p) + ig_j\tilde{\alpha}(2\omega_d-\omega_p) \left\{2 \tilde{\rho}_e(0)  -1 \right\} + i\Omega \left\{2  \tilde{\epsilon}^*_{e,j}(\omega_p-\omega_d) \right\}=0 \\
&\hspace{-0.9 cm}-[i(\omega_p-\omega_d) +\Gamma]\tilde{\epsilon}^*_{e,j}(\omega_p -\omega_d)+ ig_j\left\{\tilde{\alpha}^{*}(\omega_p)\tilde{\rho}(\omega_d) - \tilde{\alpha}(2\omega_d - \omega_p)\tilde{\rho}^{*}(\omega_d) \right\} + i\left\{\Omega^*\tilde{\epsilon}_j(2\omega_d-\omega) -\Omega\tilde{\epsilon}^*_j(\omega) \right\} = 0  
\end{align}
\end{subequations}
If the cavity is only driven at one frequency, $\omega_p$, ${\alpha}(2\omega_p - \omega) = 0$ in above equations. From~(\ref{set1}) it can be seen that in addition to oscillating at the frequency $\omega_p$ the atomic coherence also oscillates at frequencies $\omega_p \pm \Delta_{pd}$, where $\Delta_{pd} = \omega_p -\omega_d$. This was an assumption made by B. R. Mollow in his article~\cite{mollow_stimulated_1972} for solving free space scenario of the same problem as ours. We arrive at it using Fourier transforms.   
Hence, for non-zero $\tilde{\epsilon}^*_{e,j}(\omega_d-\omega)$ we get,
\begin{eqnarray}
\tilde{\epsilon}_j(\omega_p)=  -ig_j\tilde{\alpha}(\omega_p)[C_1(\omega_p)+C_2(\omega_p)]
\end{eqnarray}
Where $C_1(\omega_p)$ is same as in (\ref{firstap}) and the value of $C_2(\omega_p)$ is,
\begin{equation}
C_2(\omega_p)= \frac{-16 |\Omega|^2  (\Gamma -i\Delta_{dp}) \left[\Gamma ^2-2 i \Gamma\Delta_{dp}+4\Delta_{pa} (\Delta_{pa}-\Delta_{dp})\right]}{(\Gamma -2 i (\Delta_{pa}+\Delta_{dp})) \left(\Gamma ^2+8 |\Omega|^2+4\Delta_{pa}^2\right)\left[(\Gamma -i\Delta_{dp}) \left(4\Delta_{pa}^2+(\Gamma -2 i\Delta_{dp})^2\right)+8 |\Omega|^2(\Gamma -2 i\Delta_{dp})\right]}
\end{equation}
Finally, $\tilde{\alpha}(\omega_p)$ takes the form,
\begin{eqnarray}
 \tilde{\alpha}(\omega_p)&=&  \frac{-\eta - i \tilde{\rho}(\omega_d)\delta(\omega_d-\omega_p)\sum^{N}_{j}g_j}{\left\{\kappa- i\Delta_{pc} \right\}   + g_t^2 [C_1(\omega_p)+C_2(\omega_p)]}
 \label{alphafull}
 \end{eqnarray}

\subsection{Mollow Spectrum}\label{mollowspectrum}
\begin{figure}[H]
\centering
\includegraphics[width=7cm]{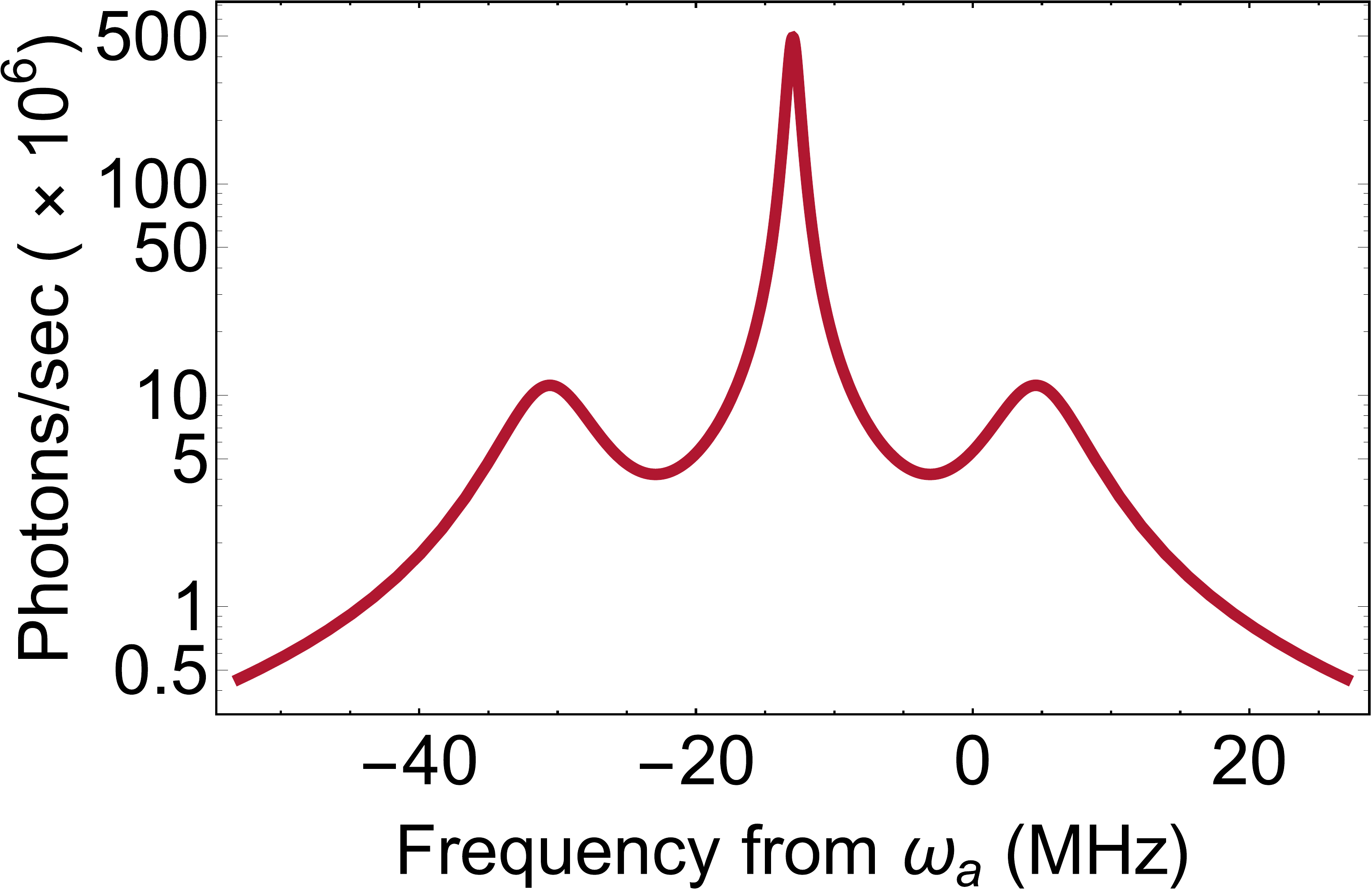}
\caption{Numerically computed Mollow spectrum on a log scale for parameters $\Omega/(2\pi) = 6$ MHz, $\Delta_{pa}/(2\pi) = -13$ MHz, $N_c = 22 \times 10^3$, $\Gamma/(2\pi) = 6.06$ and MOT laser linewidth 1 MHz. Most of the light emitted is at MOT laser/drive frequency. y-axis shows the rate of photon emission into the cavity at that particular wavelength. This rate is upper bound on the actual rate because in the calculation of the rate we assume that all the atoms are at the centre of the cavity and hence all the light emitted by the atoms which fall on the cavity mirrors come back to the centre and form the cavity mode. Also, the emission from atoms is assumed to be isotropic. Hence the rate is $\zeta \times$ total rate of photon emission, where $\zeta = 0.009$ depends on the solid angle subtended by the mirrors at the centre of the cavity. The red sideband and blue sideband have centre frequencies $-30.6$ MHz $ \approx(\Delta_{pa}-\sqrt{4\Omega^2 + \Delta_{pa}^2})/(2\pi)$ MHz and $4.6$ MHz $\approx(\Delta_{pa}+\sqrt{4\Omega^2 + \Delta_{pa}^2})/(2\pi)$ respectively.}
\label{mollow}
\end{figure}   
%

\begin{acknowledgments}
We thank G. S. Agarwal, H. Wanare and A. Narayanan for comments on the manuscript. R.S. and S.A.R. acknowledge support from the Indo-French Centre for the promotion of Advanced Research-CEFIPRA Project No. 5404-1.  
\end{acknowledgments}

\twocolumngrid
\bibliography{ref2.bib}{}
\bibliographystyle{unsrt}
\end{document}